\documentclass[a4paper,fleqn,usenatbib,onecolumn]{mnras}
\usepackage[T1]{fontenc}
\usepackage{ae,aecompl}
\usepackage{graphicx}	
\usepackage{amsmath}	
\usepackage{amssymb}	
\newcommand{\VEC}[1]{\vec {#1} }
\title[
Magnetic-field evolution with velocity circulation
]{
Magnetic-field evolution with
large-scale velocity circulation in a neutron-star crust
}
\author[Y. Kojima and K. Suzuki]{
Yasufumi Kojima\thanks{%
E-mail: ykojima-phys@hiroshima-u.ac.jp} and Kazuki Suzuki\\
Department of Physics, Hiroshima University, Higashi-Hiroshima, Hiroshima 
739-8526, Japan}
\begin{document}
 \label{firstpage}
 \pagerange{\pageref{firstpage}--\pageref{lastpage}}
  \maketitle
\begin{abstract}
  We examine the effects of plastic flow that appear in a neutron-star crust
when a magnetic stress exceeds the threshold.
The dynamics involved are described using the Navier--Stokes equation comprising the 
viscous-flow term, and the velocity fields
for the global circulation are determined using quasi-stationary approximation. 
   We simulate the magnetic-field evolution by taking into consideration
the Hall drift, Ohmic dissipation, and fluid motion induced by the Lorentz force.
The decrease in the magnetic energy is enhanced, as 
the energy converts to the bulk motion energy and heat.
It is found that the bulk velocity induced by the Lorentz force 
has a significant influence in the low-viscosity and strong-magnetic-field regimes.
This effect is crucial near magnetar surfaces.
\end{abstract}


\section{Introduction}
 The magnetic fields of neutron stars are inferred from spin-down
measurements, and the surface dipole field-strength $B_{0}$ 
ranges from $ B_{0}=10^{8}$ to $10^{15}$ G
\citep[e.g.,][for a review]{2019RPPh...82j6901E}.
Among neutron stars, highly magnetized objects are classed as magnetars
\citep{1995MNRAS.275..255T}.
They are relatively young, i.e., their characteristic ages 
are less than $10^{5}$yr; however, there exist a few exceptions
\citep[for a review]{2014ApJS..212....6O,
2015RPPh...78k6901T,2017ARA&A..55..261K}.
Magnetar activity is powered by their strong magnetic field.
Their energy sources are likely to be exhausted in the timescale 
of approximately $10^{5}$yr, which is when their activity ceases. 
This fact is an evidence of magnetic-field evolution
in a high-field-strength regime of approximately $10^{13}$G.
It is also interesting to note that the younger known radio pulsars tend to have 
stronger fields than typical known radio pulsars. 
This may suggest that the magnetic fields of radio pulsars
decay with time\citep{2014ApJS..212....6O}.
The boundary between magnetars and normal radio pulsars is unclear.
For example, SGR 0148+5729 ($B_{0}=6\times 10^{12}$G)\citep{2010Sci...330..944R}
and Swift J1822.3-1606 ($B_{0}=1.4\times 10^{13}$G) 
\citep{2012ApJ...754...27R, 2014ApJ...786...62S}
exhibit magnetar activity.
The two radio pulsars PSR J1846-0258 ($B_{0} = 4.9 \times 10^{13}$G) in 
SNR Kes 75\citep{2008Sci...319.1802G} and PSR J1119-6127
($B_{0} = 4.1\times 10^{13}$G) in SNR G292.2-0.5\citep{2016ApJ...829L..21A}
have exhibited magnetar-like transient X-ray activity.
However, PSR J1847-0130 ($B_{0} = 9.4 \times 10^{13}$G)
\citep{2003ApJ...591L.135M} has neither
shown detectable X-rays nor magnetar outbursts thus far.
Thus, it has not been discriminated by surface dipole field only.

  Recently, a young pulsar PSR J1640-4631 in SNR G338.3-0.0 has been attracting attention.
Its characteristic age is 3350 yr, $B_{0} = 1.4\times 10^{13}$G
\citep{2014ApJ...788..155G}, 
and the braking index  is $n = 3.15$\citep{2016ApJ...819L..16A}.
In the magnetic dipole-radiation model, we have $n=3$.
The majority of braking indices that have been reliably measured thus far were less than 3. 
A source with $n > 3$ is unique.
There have been are several discussions regarding the deviation from $n=3$, which include the considerations of a
change in magnetic inclination angle, 
higher multipole radiation including gravitational waves, 
and magnetic-field evolution.
The decay of the dipole magnetic field also results in $n>3$.
Detailed models have been presented to account for $n=3.15$, in which case the decay timescale of approximately $10^{5}$yr is used for phenomenological 
fitting\citep{2017ApJ...849...19G, 2019PhRvD..99h3011C}.
Such an evolution time of the dipole field 
requires theoretical foundations, or else the model might be rejected.
We expect that higher multipoles generally decay fast, and the 
dipole remains strong over a longer timescale.
Theoretical evolution models are required to be further improved.

The magnetic-field evolution in an isolated neutron star was studied 
after \cite{1992ApJ...395..250G}
discussed the involved mechanisms, but it has not yet fully understood.
Actual evaluation of the magnetic-field evolution depends on the interior state, i.e., the
relevant microphysics depend on the internal density and temperature.

In the case of magnetic fields in the core, a stronger field of up to approximately $10^{18}$G
may be confined.
The evolution of core-field is important, if the surface-field variation is closely related 
to the interior variation. The magnetic-field evolution is governed by a fluid mixture of neutrons, 
protons, and electrons.
In addition, the superfluidity in the neutron star matter should be taken into accounted. 
Therefore, the dynamics of the core-field evolution is rather complex.
This has been addressed in several studies
\citep[e.g.,][]{2011MNRAS.413.2021G,2015MNRAS.453..671G,2016MNRAS.462.1453G,
2017MNRAS.469.4979P,2017MNRAS.471..507C},
but is still being debated\citep{2017PhRvD..96j3012G,
2018MNRAS.473.4272K,2018PhRvD..98d3007O}.

The magnetic field in the crust of a neutron-star is simple.
In the crust, ions are fixed in a lattice, while only the electrons are mobile.
These dynamics may be described based on Electron Magneto-Hydro-Dynamics (EMHD).
The evolution of the magnetic field owing to the Hall drift 
and Ohmic dissipation has been studied in several  researchers.
The magnetic fields are numerically simulated 
on a secular timescale\citep[refer][for axially symmetric cases]
{2007A&A...470..303P,2012MNRAS.421.2722K,
2013MNRAS.434..123V,2013MNRAS.434.2480G,2014MNRAS.438.1618G}.
Recently, three-dimensional calculations of the evolution have been performed
\citep{2015PhRvL.114s1101W,2016PNAS..113.3944G,2018ApJ...852...21G}.
The simulation shows that non-axisymmetric features such as magnetic 
hot spots produced by an initial perturbation persist over a long 
timescale $\sim 10^6$yr.
Small-scale features are likely to decay owing to the action of  
Ohmic dissipation, but they seem to be steadily supported by the Hall effect.
This effect becomes more important as the field-strength increases.
At the same time,
magnetic stress causes deformations in the ion lattice.
The crust responds elastically when the deviation
is within a critical limit. 
Beyond this limit, the crust cracks or responds plastically.
The effect of elastic deformation on the crustal magnetic-field
evolution has been studied\citep{2018MNRAS.473.2771B}.
Sudden crust breaking could produce a magnetar outburst and/or a fast 
radio burst\citep{2016ApJ...833..189L,2018MNRAS.480.5511B,2019MNRAS.488.5887S}.
Plastic flow beyond the critical point is crucial for 
long-term evolution\citep{2016ApJ...824L..21L},
and a coupled system between the flow and magnetic field can be solved numerically
in a two-dimensional square box of a crust-depth size\citep{2019MNRAS.486.4130L}.
Their simulation shows significant motion near the surface
at a rate of 10--100 cm yr$^{-1}$ for a certain viscosity range.
At $10^{3}$yr, the path length reaches $\sim 1$km, 
which is the size of the simulation in plane-parallel geometry.
The magnetic-field structure comprises a poloidal loop field confined in a two-dimensional box.
It is important to examine the effect of the plastic
flow on the global features.
In particular, changes in the external dipole field 
in $10^{3}$--$10^{5}$yr are interesting.

It is not trivial to calculate the flow over a long span with respect
to space and time.
The plastic flow is represented by a viscous-flow term in 
the equation of motion, and the corresponding dynamics is described by the 
standard Navier--Stokes equation.
It is impossible to perform a direct time integration of the equation, and thus, we adopt several approximations.
The flow is assumed to be circulative,
although the Lorentz force induces two types of vectorial deformation:
solenoidal and irrotational vector fields.
The Lorentz force has a magnitude of the order 
of $10^{-4}(B/10^{15}{\rm G})^2$ to dominant forces, i.e., 
gravity and degenerate pressure, such that stellar motions driven 
by the Lorentz force are highly restricted.
A solenoidal motion does not accompany any change in density 
distribution, on which the dominant forces depend.
Therefore, such a flow pattern may be extended to a large scale.
In contrast, irrotational motion is highly suppressed in a star.
We also assume that all the forces are instantaneously balanced, as
our relevant timescale of secular evolution is much larger.
A sequence of equilibria is calculated for the fluid motion.
We simplify the magnetic-field configuration as
our concern is the global feature. 
As our initial model, we consider the ordered magnetic fields 
to eliminate the short-timescale events associated 
with a small-scale structure.
The fields are always considered to exist from the crust to the exterior to 
avoid the effects of the core-field evolution.
During the evolution, rearrangement of the magnetic field and flows 
associated with a burst/flare may occur in the dynamical timescale.
Such a rearrangement is less frequent at a larger scale.
Moreover, the physics required to determine a new state is very complicated in nature.
Sudden rearrangement of magnetic field is ignored in our simulation.

We simulate the magnetic-field evolution coupled
to the global viscous-fluid motion, which is assumed to start 
as a plastic flow in the entire crust.
Models and equations relevant to our problems are discussed 
in Section 2. 
The numerical results of a long-term  magnetic-field evolution are presented in Section 3.
Finally, our conclusions are presented in Section 4.

\section{Formulation and model}
  The evolution of the magnetic field  ${\VEC B}$ is governed by 
the induction equation,
\begin{align}
\frac{\partial}{\partial t}{\VEC B}=-{\VEC \nabla}\times (c{\VEC E}).
    \label{Frad.eqn}
\end{align}
The electric field ${\VEC E}$ is given by the generalized Ohm's law,
\begin{equation}
c{\VEC E}=\frac{c}{\sigma}{\VEC j} 
+\frac{1}{{\rm e} n_{\rm e}}{\VEC j}\times {\VEC B}
-{\VEC v}_{\rm b}\times {\VEC B}
 \equiv\frac{c}{\sigma}{\VEC j}-{\VEC v}\times{\VEC B},
    \label{Edef.eqn}
\end{equation}
where
\begin{equation}
 {\VEC v}\equiv{\VEC v}_{\rm b}-\frac{1}{{\rm e} n_{\rm e}}{\VEC j}.
    \label{vsum.eqn}
\end{equation}
In eq.(\ref{Edef.eqn}), $\sigma$ represents the electric conductivity, 
$ n_{\rm e}$ is the electron number density, and the velocity 
${\VEC v}_{\rm b}$ represents the bulk flow velocity.
The velocity may be approximated as the ion velocity owing to
a large mass difference between the ions and electrons. 
In a co-moving frame of a neutron star, the ions in the crust are fixed 
in the lattice, and thus, we have ${\VEC v}_{\rm b}=0$.
In this case, the dynamics of eq.(\ref{Frad.eqn})
is completely determined by the magnetic fields only as
the  electric current ${\VEC j}$ is related to 
the magnetic field as per Amp{\'e}re--Bio--Savart's law,
\begin{equation}
  {\VEC j}=\frac{c}{4\pi}{\VEC \nabla}\times {\VEC B}.
 \end{equation}
This problem has been thus so far studied from various viewpoints
\citep[e.g.,][]{2007A&A...470..303P,2012MNRAS.421.2722K,
2013MNRAS.434..123V,2013MNRAS.434.2480G,2014MNRAS.438.1618G,
2015PhRvL.114s1101W,2016PNAS..113.3944G,2018ApJ...852...21G}.

When a magnetized neutron star is born, all the forces are settled and balanced within a short timescale.
The Lorentz force ${\VEC j}\times {\VEC B}$
changes in a secular timescale according to the magnetic-field evolution.
Therefore, the balance of the forces eventually breaks down.
The structure change may comprise a slow or abrupt process 
depending on the material property.
When a magnetic stress exceeds a threshold, the crust 
may be completely fractured, and the magnetic field is rearranged.
In this paper, we take into consideration the opposite case.
Thus, we treat the gradual deformation motion of the ions of the lattice as
a plastic flow and examine its effect on the magnetic-field evolution,
i.e., the effect of a new velocity component ${\VEC v}_{\rm b}$, as 
shown in eq.(\ref{Edef.eqn}).

\subsection{Quasi-stationary approximation}
  The equation of bulk flow motion is given by
\begin{align}
\rho\left(\frac{\partial}{\partial t}+{\VEC v}_{\rm b}\cdot{\VEC \nabla}\right) {\VEC  v}_{\rm b}={\VEC f},
   \label{eqnavier}
\end{align}
where $\rho$ represents the mass density, and ${\VEC f}$ is the force per unit volume.
The force density ${\VEC f}$ in the crust comprises pressure, gravity, 
the Lorentz force, and the plastic-flow term. 
The deformation is elastic 
below a threshold, but we ignore the regime in this paper.
We assume an incompressible fluid for the sake of simplicity and explicitly consider
\begin{equation}
{\VEC f}=-{\VEC \nabla}P-\rho{\VEC \nabla}\Phi_{\rm G}
+\frac{1}{c}{\VEC j}\times {\VEC  B}+\nu\nabla^{2}{\VEC  v}_{\rm b} ,
   \label{Navier.eqn}
\end{equation}
where $\nu $ is the viscosity coefficient. The last term in 
eq.(\ref{Navier.eqn}) is physically the same as a 
standard viscous-flow term \citep{2016ApJ...824L..21L}.
Based on the condition ${\VEC \nabla} \cdot {\VEC v}_{\rm b}=0$,
the last term may be written as
\begin{equation}
\nu\nabla^{2}{\VEC v}_{\rm b}=-\nu{\VEC \nabla}\times ({\VEC \nabla}\times {\VEC v}_{\rm b}) .
\end{equation}

It is impossible to simultaneously follow the bulk flow motion given by eq.(\ref{eqnavier}),
when the long-term evolution of the magnetic fields is calculated. 
The order of their timescales is extremely different.
As our relevant timescale is much longer than the dynamical timescale,
we approximate ${\VEC f}=0$.
That is, the forces are always balanced for a short timescale,
such that the velocity-field structure is 
subjected to an equilibrium state in a moment.

Another difficult problem arises in calculating the 
velocity field based on the force balance ${\VEC f}=0$.
There are four types of forces involved in eq.(\ref{Navier.eqn}), but 
their magnitudes are quite different.
The degenerate pressure and gravity represented by eq.(\ref{Navier.eqn})
are dominant terms, and they actually determine the stellar structure.
The barotropic structure is generally a good approximation.
Thus, the first and second terms are regarded as irrotational vectors. 
It should be noted that any vector can be decomposed by using an irrotational vector and a solenoidal vector
 \citep[Helmholtz decomposition, see][]{2005mmp..book.....A}.
The Lorentz force comprises a mixture of the irrotational and solenoidal vectors.
The solenoidal part produced during the magnetic-field evolution
has no relation to the dominant forces, i.e., pressure and gravity.
We herein assume that the bulk motion arises from the
solenoidal part of the Lorentz force.
Thus, a small component of the magnitude is extracted by the vectorial 
property. The velocity ${\VEC v}_{\rm b}$ is assumed to be determined by
\begin{equation}
 {\VEC \nabla}\times {\VEC f}={\VEC \nabla}\times 
\left[\frac{1}{c}{\VEC j}\times {\VEC B}+\nu\nabla^{2}{\VEC  v}_{\rm b}\right]=0.
    \label{assumption1}
\end{equation}

Our concern in terms of this paper is how a new component 
${\VEC v}_{\rm b}$ in eq.(\ref{Edef.eqn}) 
 will change the magnetic-field evolution given by eq.(\ref{Frad.eqn}).
We explicitly describe how the bulk velocity ${\VEC v}_{\rm b}$ is 
determined in an axially symmetric case.
A general form of the incompressible flow is given by two functions $F$ and $H$
\begin{equation}
{\VEC v}_{\rm b}={\VEC \nabla}\times 
\left(\frac{F}{\varpi}{\VEC e}_{\phi}\right)+\frac{H}{\varpi}{\VEC e}_{\phi},
 \label{dfvel.eqn}
\end{equation}
where ${\VEC e}_{\phi}$ is a unit vector in the $\phi$
direction, and $\varpi=r\sin \theta$ in spherical 
coordinates $(r,\theta,\phi)$.
The vorticity is calculated as
\begin{equation}
{\VEC \nabla}\times {\VEC v}_{\rm b}= 
 {\VEC \nabla}\times \left(\frac{H}{\varpi}{\VEC e}_{\phi}\right) 
+\frac{W}{\varpi}{\VEC e}_{\phi}.
 \label{vort.eqn}
\end{equation}
Here, a new function $W$ is introduced by $F$,
\begin{equation}
\left[ {\VEC \nabla} \times {\VEC \nabla} \times \left(
\frac{F}{\varpi}{\VEC e}_{\phi} \right)  \right]_{\phi}=\frac{W}{\varpi}.
 \label{ff2nd.eqn}
\end{equation}
This equation is explicitly written as
\begin{equation}
\frac{\partial^{2} F}{\partial r^{2}}
+\frac{\sin\theta}{r^{2}}\frac{\partial}{\partial\theta}
\left(\frac{1}{\sin\theta}\frac{\partial F}{\partial\theta} \right)=-W.
\end{equation}
Based on the assumption of the axially symmetric configuration, 
$f_\phi=0$ is reduced to
\begin{equation}
\left[{\VEC \nabla}\times {\VEC \nabla}\times 
\left(\frac{H}{\varpi}{\VEC e}_{\phi} \right)\right]_{\phi} 
=\frac{1}{c\nu}({\VEC j}\times {\VEC B})_{\phi}.
 \label{hh2nd.eqn}
\end{equation}
Equation (\ref{assumption1}) for $W$ is reduced to
\begin{equation}
 \left[{\VEC \nabla} \times \left(c\nu{\VEC \nabla} \times 
\frac{W}{\varpi} {\VEC e}_{\phi} \right)\right]_{\phi}= \varpi\left[
{\VEC B} \cdot {\VEC \nabla}\left(\frac{j_{\phi}}{\varpi} \right) 
 -{\VEC j} \cdot {\VEC \nabla}\left(\frac{B_{\phi}}{\varpi} \right)\right] ,
  \label{ww2nd.eqn}
\end{equation}
where the conditions 
${\VEC \nabla}\cdot{\VEC j}={\VEC \nabla}\cdot{\VEC B}=0$ are used.
Thus, the bulk velocity field is determined 
by solving two second-order partial differential equations
(\ref{hh2nd.eqn}) and (\ref{ww2nd.eqn})
of an elliptic type, and the stream function $F$ of the velocity 
in eq.(\ref{dfvel.eqn}) is determined by solving eq.(\ref{ff2nd.eqn}).

\subsection{Evolution of magnetic field}
Numerical methods for describing the magnetic-field evolution 
are discussed\citep[][for a review and references]{2019LRCA....5....3P}.
There are several choices of variables, but 
we calculate the time-integration of $B_{\phi}$ and $j_{\phi}$
in this study. From eq.(\ref{Frad.eqn}) and on applying a rotation to it, 
we obtain a set of evolution equations for $B_{\phi}$ and $j_{\phi}$
\begin{align}
\frac{\partial}{\partial t}B_{\phi}
=-\left[{\VEC \nabla}\times
\left(\frac{c^2}{4\pi\sigma}{\VEC \nabla}\times (B_{\phi}{\VEC e}_{\phi}) 
-{\VEC v}\times {\VEC B} \right)\right]_{\phi},
\end{align}
and
\begin{align}
\frac{\partial}{\partial t}\left(\frac{4\pi j_{\phi}}{c}\right)
=-\left[{\VEC \nabla}\times {\VEC \nabla}\times 
\left(cE_{\phi}{\VEC e}_{\phi} \right)\right]_{\phi} .
    \label{evlj3.eqn}
\end{align}

Once $B_{\phi}$ and $j_{\phi}$ are integrated with respect to time, 
the poloidal current ${\VEC j}_{\rm p}$ 
and poloidal magnetic field ${\VEC B}_{\rm p}$ 
at each time are determined using 
\begin{equation}
\frac{4\pi{\VEC j}_{\rm p}}{c} 
={\VEC \nabla} \times \left(B_{\phi}{\VEC  e}_{\phi} \right),
  \label{Amppol.eqn}
\end{equation}
\begin{equation}
\left[{\VEC \nabla}\times {\VEC B}_{\rm p}\right]_{\phi}
=\frac{4\pi j_{\phi}}{c}.
  \label{amp3.eqn}
\end{equation}
It is convenient to solve the last equation by introducing 
a magnetic function $\Psi$ that satisfies
\begin{equation}
{\VEC B}_{\rm p}={\VEC \nabla}\times \left(
\frac{\Psi}{\varpi}{\VEC e}_{\phi} \right).
 \label{BpPhi.eqn}
\end{equation}
Equation (\ref{amp3.eqn}) is an elliptical equation of $\Psi$,
\begin{equation}
\left[{\VEC \nabla}\times {\VEC \nabla}\times 
 \left(\frac{\Psi}{\varpi}{\VEC e}_{\phi}
 \right)\right]_{\phi}=\frac{4\pi j_{\phi}}{c}.
  \label{G2nd.eqn}
\end{equation}
  Instead of solving eq.(\ref{G2nd.eqn}) at each time step, 
another method comprises the time-integration of $\Psi$.
From eqs.(\ref{Frad.eqn}) and (\ref{evlj3.eqn}),
the evolution of $\Psi$ is given by the azimuthal component of the electric field,
\begin{equation}
\frac{\partial \Psi}{\partial t}=-c\varpi E_{\phi}.
 \label{evlPsi.eqn}
\end{equation}
We adopt eq.(\ref{evlPsi.eqn}) as a numerical method, as
it is found to be accurate and fast in our numerical calculations.

It is instructive to write down energy-conservation equation\footnote{%
The meridional flow will have associated changes in chemical composition and 
may thus trigger e.g. electron capture reactions that might generate entropy. 
However, these are likely points for future research.
We here consider the magnetic-field evolution in a fixed structure.}.
The magnetic energy inside a crust of volume $V$ is given by
\begin{equation}
 E_{\rm B}=\frac{1}{8\pi}\int_{V}B^2 dV, 
\end{equation}
and its time derivative is given by
\begin{equation}
\frac{dE_{\rm B}}{dt}=-\int_{V}\frac{j^2}{\sigma}dV
-\frac{1}{c}\int_{V} ({\VEC j}\times {\VEC B} )\cdot{\VEC v}dV
-\frac{c}{4 \pi}\int_S 
({\VEC E }\times {\VEC B })\cdot d{\VEC S }.
 \label{Ebalance.eqn}
\end{equation}
The first term represents the Joule loss, while the second represents the work done 
by the Lorentz force inside the crust. 
For the Hall evolution, i.e., the velocity in eq.(\ref{vsum.eqn}) is 
given by ${\VEC v}=-{\VEC j}/({\rm e} n_{\rm e})\equiv {\VEC v}_{\rm H}$
such that the work vanishes. 
The third term in eq.(\ref{Ebalance.eqn})
represents the Poynting energy lost through a surface. 
We found that this term is very small in our 
secular simulation of magnetic fields.

\subsection{Boundary conditions}
 We solve a set of time-dependent equations for $B_{\phi}$, $j_{\phi}$,
and $\Psi$, and a set of elliptic-type equations for the bulk velocity fields
$W$, $F$, and $H$ inside a crust. The spherical shell region of the crust 
is represented by $r_{c} \le r \le R$ and $0 \le \theta \le \pi$.
We discuss the boundary conditions for these functions.

The boundary condition at the symmetric axis
($\theta =0, \pi$) is a regularity condition.
That is, all the functions $B_{\phi}$, $j_{\phi}$, $\Psi$, $W$, 
$F$, and $H$ should vanish at the axis.

  We assume that the magnetic fields are expelled from the inner core 
and that they are maintained by electric currents flowing in the crust.
This assumption simplifies the problem, as core magnetic fields 
do not affect crustal fields. We impose the conditions
$B_{\phi}=j_{\phi}=\Psi=0$ for the magnetic fields. 
Based on eqs.(\ref{Amppol.eqn}) and (\ref{BpPhi.eqn}), these conditions indicate that 
no radial components of the current and magnetic field cross 
the boundary at $r_{c}$.
In the case of the velocities, we impose the conditions $W=F=H=0$, that is,
the bulk flow is suppressed at the interface of the core.

   Finally, we discuss the conditions at the surface $r=R$.
The toroidal magnetic field and electric current are confined inside a star, 
such that we impose $B_{\phi}=j_{\phi}=0$.
The poloidal magnetic field is not confined inside the star
but is continuously connected to the exterior vacuum solution.
We impose the continuity of the magnetic function $\Psi$ 
as the boundary condition.
This indicates that the continuity of $B_r$ and discontinuity of 
a tangential component are generally allowed by the surface current.
In the case of the bulk motions, the fixed boundary 
conditions $W=F=H=0$ are imposed. These are the same as those at $r_{c}$.

\subsection{Crust model}
 Our consideration is limited to the inner crust of a neutron star,
where the mass density ranges from 
$\rho_{c} =1.4\times 10^{14}$ g cm$^{-3}$ 
at the core-crust boundary $r_{c}$ 
to $\rho_{1} = 4\times 10^{11}$ g cm$^{-3}$
at the neutron-drip point $R$.
We ignore the outer crust, the structure of which
may be significantly affected in the presence of a strong magnetic field, and
we treat the exterior region of $r> R$ as the vacuum.
The crust thickness is assumed to be $\Delta r/R=(R-r_{c})/R=0.1$.

The density profile of $r_c \le r \le R$
is approximated as follows \citep{2019MNRAS.486.4130L}.
\begin{equation}
{\hat \rho}\equiv\frac{\rho}{\rho_{c}}
=\left[1-\left(1-\left(\frac{\rho_1}{\rho_{c}}\right)^{1/2}
 \right)\left(\frac{r-r_{c}}{\Delta r} \right)\right]^2 .
 \end{equation}
Using the normalized density ${\hat \rho}$, the distribution of 
the electron number density $n_{\rm e}$, electric conductivity $\sigma$, and
viscous coefficient $\nu$ are approximated using 
analytical expressions\citep{2019MNRAS.486.4130L}.
 \begin{align}
\label{ne_express}
& n_{\rm e}  =n_{\rm e c}
\left( 0.44{\hat \rho}^{2/3} + 0.56{\hat \rho}^{2}\right), \\
\label{sigm_express}
& \sigma=\sigma_{c} 
\left(0.44{\hat \rho}^{2/3} +0.56{\hat \rho}^{2}\right)^{2/3},  
\\
\label{nuvis_express}
& \nu=\nu_{\rm c}\left(0.44{\hat \rho}+0.56{\hat \rho}^{3}\right),
 \end{align}
The numerical coefficients in front of these equations are
 \begin{align}
& n_{\rm e c}=3.4\times 10^{36}{\rm cm }^{-3}, \\
& \sigma_{\rm c}=1.5\times 10^{24}{\rm s}^{-1}, \\
& \nu_{c}=1.4\times 10^{38}{\rm g cm }^{-1}{\rm s}^{-1} .
 \label{nuc.coefficient}
\end{align}
We consider a simple form of electric conductivity 
$\sigma\propto n_{\rm e}^{3/2}$, but the coefficient $\sigma_{c}$ may
change by an order of magnitude in the actual situation,
as $\sigma_{c}$ depends on the thermal evolution of the neutron star.
Using these numerical values,
we estimate the typical timescales, the Ohmic 
timescale $\tau_{\rm Ohm}$, Hall timescale $\tau_{\rm H}$, and
bulk motion timescale $\tau_{\rm v}$ as follows.
\begin{align}
& \tau_{\rm Ohm}=\frac{4\pi \sigma_{\rm c} \Delta r ^2}{c^2}
  =5.7\times 10^6{\rm yr}\left(\frac{\Delta r}{1{\rm km}} \right)^2,
   \label{timescale.ohm}
\\
& \tau_{\rm H} =\frac{4\pi {\rm e} n_{\rm e c} \Delta r ^2}{c B_{\rm av}}
=6.9\times 10^5{\rm yr}
\left(\frac{ B_{\rm av} }{10^{14.5}{\rm G} }\right)^{-1} 
\left(\frac{\Delta r}{1{\rm km}} \right)^2, 
   \label{timescale.hall}
\\
& \tau_{\rm v} =\frac{4\pi \nu_{\rm c}}{ B_{\rm av} ^2} 
=5.5\times 10^2{\rm yr}
\left(\frac{B_{\rm av}}{10^{14.5}{\rm G} }\right)^{-2} .
   \label{timescale.tv}
\end{align}
Herein, we used the maximum values for $n_{\rm e}$, $\sigma$, and $\nu$,
i.e., the values at the core--crust boundary. 
The magnetic field in our model is zero at the boundary, such that
the volume averaged value $B_{\rm av}$ is used in these estimates.
It should be noted that the actual timescales vary substantially. 
In particular, $n_{\rm e}$ decreases by a factor of $10^{-2}$
and $\nu$ by a factor of $10^{-3}$ at the surface $r=R$.
The configuration near the surface changes in much smaller timescales
than the above ones.
The timescale $\tau _{\rm v}$ is relevant to our quasi-stationary approximation.
In a certain timescale $\ll \tau_{\rm v}$, all the 
forces are assumed to be balanced, and the bulk velocity is determined.
When $\tau_{\rm v}$ becomes too small for some parameters, 
the quasi-stationary approximation may not be justified, and
we have to take into consideration microscopic dynamics.

  The following relation by order-of-magnitude may be useful for understanding 
the effect of the bulk motion in the numerical simulation.
\begin{equation}
\frac{v_{b}}{v_{\rm H}}\approx
\frac{(\nu c)^{-1} jB \Delta r^{2}}{({\rm e} n_{\rm e})^{-1}j}
\approx\frac{\tau_{\rm H}}{\tau _{\rm v}}\propto \nu^{-1}B,
   \label{ratiothtv}
 \end{equation}
where $v_{\rm H}$ is the Hall velocity that is given by the second term in eq.(\ref{vsum.eqn}).
Equation (\ref{ratiothtv}) indicates that
the bulk velocity $v_{\rm b}$ increases as the viscosity decreases.
It should also be noted that both  $v_{\rm b}$ and $v_{\rm H}$ increase
as the magnetic field strength increases, but
the ratio $v_{\rm b}/v_{\rm H}$ increases.
Thus, the bulk velocity induced by the Lorentz force is 
important in low-viscosity and strong-magnetic-field regimes.

\section{Evolution}
\subsection{Initial model}
  We discuss the initial model of the magnetic fields.
A reasonable state comprises a barotropic MHD equilibrium
\citep{2013MNRAS.434.2480G}. 
We assume that the toroidal field vanishes, i.e., $B_{\phi}=0$.
The azimuthal component of the electric current $j_{\phi}$ for 
the barotropic MHD equilibrium is given as follows\citep{2013MNRAS.434.2480G}. 
\begin{equation}
 j_{\phi}=\rho f_{\rm B}r\sin \theta,  
 \label{simpl1.eqn}
\end{equation}
where $f_{\rm B}$ is generally a function of $\Psi$ only, and 
$f_{\rm B}$ is assumed to be a constant for the sake of simplicity.
Based on eqs.(\ref{BpPhi.eqn}) and (\ref{G2nd.eqn}), the magnetic function 
$\Psi$ and poloidal magnetic field ${\VEC B}_{\rm p}$ are calculated.
The functional form (\ref{simpl1.eqn}) with a constant $f_{\rm B}$ indicates that  
the poloidal field is given by the dipole component $l=1$ only
when we expand $\Psi$ with Legendre polynomials as
$\Psi=-\sum g_{l}\sin\theta\partial P_{l}/\partial \theta$.

The magnetic fields in the MHD equilibrium are not fixed in a longer timescale.
Their state is not in equilibrium for the electrons and is therefore evolved.
In order to explain this, we neglect the Joule term and bulk motion in 
eqs.(\ref{Frad.eqn}) and (\ref{Edef.eqn}).
The initial Lorentz force for eq.(\ref{simpl1.eqn})
is given by 
${\VEC j}\times {\VEC B}=j_{\phi}{\VEC \nabla}\Psi/\varpi$
$=\rho f_{\rm B}{\VEC \nabla}\Psi$
and the evolution of the magnetic field at the beginning is given by
\begin{equation}
\frac{\partial {\VEC B }}{\partial t}={\VEC \nabla}\Psi\times{\VEC \nabla}\chi,  
 \label{initialevl.eqn}
\end{equation}
where $\chi\equiv\rho f_{\rm B}/({\rm e} n_{\rm e})$ is
proportional to the inverse of the electron number fraction.
In our model given by eq.(\ref{ne_express}),
$\chi$ depends on the radial position, and hence, 
the magnetic field evolution is driven by the Hall term.

\subsection{Two equilibrium models}
\begin{figure}
\begin{center}
  \includegraphics[scale=1.0]{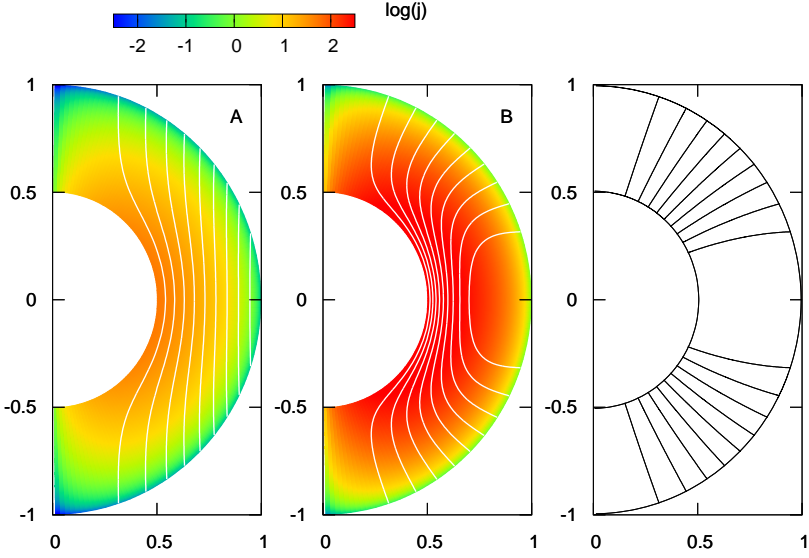}%
\caption{ 
\label{Fig1}
Magnetic function $\Psi$ indicated by contour lines and
azimuthal current $4\pi j_{\phi}R/(cB_{0})$ indicated by colors.
It should be noted that the crust region ($0.9\le r/R\le 1$) 
is enlarged by five times for display.
Left and middle panels show 
poloidal magnetic field produced by the azimuthal current. 
For comparison, the contour lines of a magnetic dipole located at the origin
are shown in the right panel.
}
\end{center}
\end{figure}

In Fig.~\ref{Fig1}, we present two models in MHD equilibrium. 
In order to display the detailed structure in the crust, 
the region $(r\sin\theta, r\cos\theta)$, ($0.9\le r/R \le 1$) 
is enlarged by five times in the figure as 
$(\xi(r)\sin\theta, \xi(r)\cos\theta)$,
where $\xi=1+(R-r)/(2\Delta r)$.
The magnetic field in both the models is matched at the surface to the
external dipole field of the same strength, but 
there is a difference in the surface current distribution at $r=R$. 
In model A shown in the left panel of Fig.~\ref{Fig1}, 
the surface current is allowed, such that $B_{\theta}$ is discontinuous.
In model B shown in the middle panel of Fig.~\ref{Fig1}, 
there is no surface current, such that
$B_{\theta}$ is continuously connected to the external field.

As shown in Fig.~\ref{Fig1}, there is a stronger current flowing 
in model B than that in model A. As a result,
the internal field of model B is stronger than that of model A. 
In order to compare their field strength quantitatively, we define 
a volume-average of the magnetic field using the root-mean-square value.
\begin{equation}
 B_{\rm av}\equiv\left(\frac{\int B^2 dV}{\int dV} \right)^{1/2} .
\end{equation}
The numerical calculation shows that the average strength is
$B_{\rm av}=4.3B_{0}$ for model A
and $B_{\rm av}=6.7B_{0}$ for model B,
where $B_{0}$ is the magnetic-field strength at the pole $(r=R, \theta =0)$.
The magnetic energy  $E_{\rm B}$ stored in the crust part
is  $E_{\rm B}=0.85B_{0}^2 R^3$ for model A
and $E_{\rm B}=2.1B_{0}^2 R^3$ for model B.
Almost twice the energy of model A is deposited in model B by the strong currents.

\subsection{Hall evolution with Ohmic dissipation}
\begin{figure}
\begin{center}
  \includegraphics[scale=1.0]{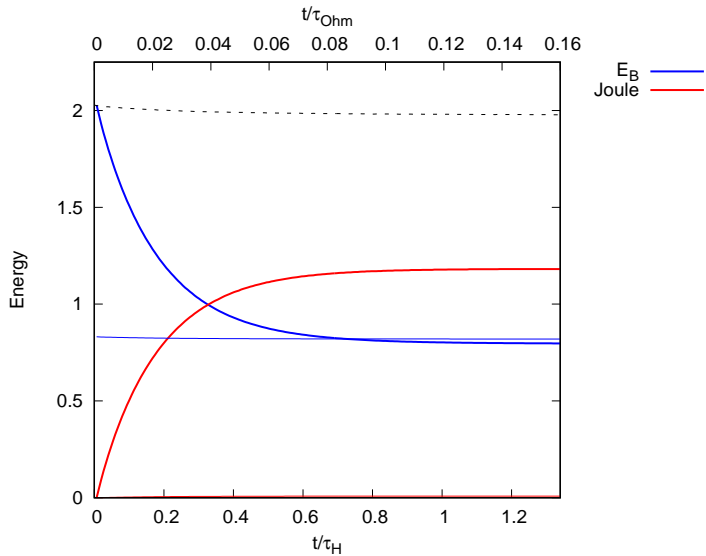}
\caption{ 
\label{Fig2}
Decay of magnetic energy.
Magnetic energy $E_{\rm B}/(B_{0}^2R^3)$ 
is represented by a decreasing curve and the time-integrated Joule-loss 
by an increasing curve.
The results for initial model B are indicated by thick lines, whereas those for 
model A are indicated by thin lines, but their variations are slow.
The horizontal dotted line denotes the numerical energy conservation for model B. 
}
\end{center}
\end{figure}

  As discussed in subsection 3.1, the magnetic fields in the MHD equilibrium 
are not fixed, but change according to the Hall drift.
The magnetic energy decays by the Joule loss on a secular timescale.
We calculated the time evolution for two initial models described in 
the previous subsection and compared the results as shown in Fig.~\ref{Fig2}.
The numerical energy conservation for the initial model B is 
also indicated by a dotted line. The sum of the magnetic energy
and time-integrated Joule-loss is almost constant.
The results for model B are indicated by thick curves.
The magnetic energy $E_{\rm B}$ decreases by half at the time 
$\sim 0.3\tau_{\rm H}$.
The results for model A are also indicated by thin lines, 
but their variations are quite slow.
The decrease is given by $\Delta E_{\rm B} =0.01B_0^2 R^3$ at 
$\tau_{\rm H}\sim 0.12\tau_{\rm Ohm}$. 
It takes a longer time $\sim \tau_{\rm Ohm}$ to observe
an evident magnetic decay in model A.
In model B, a larger amount of current is initially 
confined to the inner region and
then drifts to the outer high-resistivity region and dissipates 
in a shorter timescale $\sim 0.3 \tau_{\rm H} \sim 0.04 \tau_{\rm Ohm}$.

The numerical simulation with the Joule and Hall terms
is characterized by the so-called magnetic Reynolds number or
a magnetization parameter, $B\sigma/({\rm e} c n_{\rm e})$,
which is a ratio of the Joule to Hall timescales (see eqs.
(\ref{timescale.ohm})--(\ref{timescale.hall}).)
As an overall normalization of Fig.~\ref{Fig2},  we take 
$B_{0} =5\times 10^{13}$G ($\sim  B_{\rm Q}\equiv 4.4\times 10^{13}$G),
which is a marginal value between that of magnetars and pulsars.
The average strength in the interior for model B
is $B_{\rm av}\approx 3\times 10^{14}$G, the initial magnetic energy is
$E_{\rm B}\approx 6\times 10^{45}$erg, and its
half-life corresponds to
$0.3 \tau_{\rm H}\sim 0.04\tau_{\rm Ohm}\sim 2\times 10^{5}$yr.
It is evident that the magnetic energy increases $E_{\rm B}\propto B_{0}^2$
and the evolution timescale $\tau_{\rm H}\propto B_{0}^{-1}$ becomes short
for magnetars with $B_{0}> 5\times 10^{13}$G.
Furthermore, the dissipation rate 
($\propto  E_{\rm B} \tau_{\rm H}^{-1}\propto B_{0}^3$)
significantly depends on the magnetic field strength $B_{0}$.

The magnetic energy decreases in a timescale $\sim \tau_{\rm H}$, but 
the external field is retained as a dipole with the initial field strength.
This is related to the choice of the initial model.
We adopt a smooth configuration, in which 
the poloidal field is dipolar and toroidal field is absent.
The coupling between  the poloidal and toroidal fields is suppressed,
although the toroidal field is generally produced.
In the next subsection, we incorporate a viscous bulk flow
in this simple system, and the resulting effects are easily understood.

 \subsection{Bulk motion induced by Lorentz force}
\begin{figure}
\begin{center}
  \includegraphics[scale=1.0]{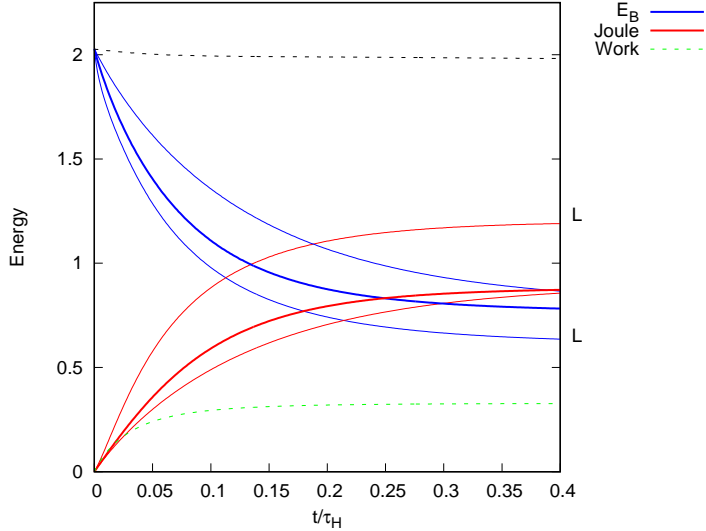}%
\caption{ 
\label{Fig3}
Decay of magnetic energy in models with bulk motion.
For our fiducial model with $\nu_{c}$,
the magnetic energy $E_{\rm B}$ is indicated by a thick decreasing curve,
the time-integrated Joule loss by a thick increasing curve,
and the work by a dotted curve.
Horizontal dotted line denotes the sum of all three.
The results of different viscosity coefficients are also presented
for the magnetic energy and time-integrated Joule loss only.
The curve with the label 'L' corresponds to the low-viscosity model.
}
\end{center}
\end{figure}

  Figure \ref{Fig3} shows the magnetic energy evolution with bulk motion
for the initial magnetic field given by model B.
On taking into consideration the uncertainty of the viscosity coefficient,
we calculated three models
by changing $\nu$ while using the same spatial profile.
The fiducial value $\nu_{c}$ is given by eq.(\ref{nuc.coefficient}), and
the magnitude is changed by a factor of 5, i.e.,
models with $0.2\nu_{c}$,  $\nu_{c}$, and $5\nu_{c}$ are calculated.

On  including the bulk motions, 
the magnetic energy is transferred to two channels: 
heat via Joule loss and bulk kinetic energy via work 
(see eq.(\ref{Ebalance.eqn})).
The latter initially acts as the magnetic energy loss, as
the bulk velocity is set as zero.
Thus, the initial decrease in the magnetic energy increases in speed owing to the bulk motion.
Half the magnetic energy is dissipated at the time $0.15 \tau_{\rm H}$ 
in the fiducial model.
The Joule loss and work rate are comparable in magnitude
at the early stage $\sim 0.05\tau_{\rm H}$.
Both of these tend to be ineffective at approximately $0.3\tau_{\rm H}$.
Till this time, the magnetic energy decrease is 
$\Delta E_{\rm B}=1.22B_{0}^2R^3$, the integrated Joule loss energy is
$0.88B_{0}^2 R^3$, and the net work is  $0.34B_{0}^2 R^3$.
On comparing these with results obtained without considering bulk motion,
it is determined that a larger amount of magnetic energy is lost  
by $\sim$4\% in the fiducial model. 
On reducing the viscosity, this effect is enhanced, and 
the value becomes 16\% in the low-viscosity model with $0.2\nu_{c}$.
The energy loss is enhanced not by the Joule heating but by the work performed.
That is, the magnetic energy is used to drive the bulk motions.

\begin{figure}
\begin{center}
  \includegraphics[scale=1.0]{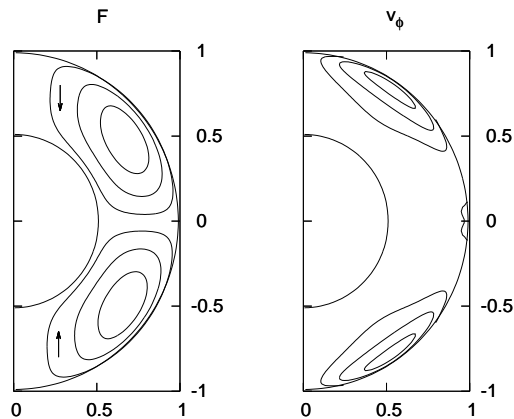}
\caption{ 
\label{Fig4}
Contours of stream function $F$(left panel) and
azimuthal velocity $v_{\phi}$(right panel)
at $0.02\tau_{\rm H}$.
The stream function $F$ is negative for $0<\theta<\pi/2$,
but positive for $\pi/2<\theta<\pi$.
An arrow in the left figure indicates the direction of 
the meridional velocity.
There are sharp maxima of $v_{\phi}$ 
in the vicinity of the surface.
There is a small region of $v_{\phi}<0$
near an equator in the right panel. 
}
\end{center}
\end{figure}

Figure \ref{Fig4} shows the fluid motion for the fiducial viscosity
at the time $0.02\tau_{\rm H}$.
In the left panel, the contours of a stream function $F$ are presented, while 
those of the azimuthal velocity $v_{\phi}$  are shown in terms of an ``enlarged" coordinate.
The stream function $F$ is anti-symmetric with respect to $\theta=\pi/2$,
that is, $F<0$ in the upper region ($0<\theta<\pi/2$), but
$F>0$ in  the lower region ($\pi/2<\theta<\pi$) in Fig.~\ref{Fig4}.
Thus, the meridional circulation flow is counter-clockwise in the upper region, 
while the flow becomes clockwise.
The maximum of $|v_\theta|$ occurs at 
$\theta\approx 0.2\pi$ and $\theta\approx 0.8\pi$
in the vicinity of the surface:
$v_\theta\approx -7$cm yr$^{-1}$ at $\theta\approx 0.2\pi$  and
$v_\theta\approx +7$cm yr$^{-1}$ at $\theta\approx 0.8\pi$.
The radial component $|v_r|$ is 
one order of magnitude smaller than $|v_\theta|$.
Using eq.(\ref{dfvel.eqn}), we have a relation between them,
$|v_r| =|\varpi^{-1}\nabla_{\theta}F|$
$\ll |v_\theta|=|\varpi^{-1}\nabla_{r}F|$,
wherein the inequality can be attributed to a steep slope of $F$ 
in the radial direction,
as inferred from the contours of $F$ in Fig.~\ref{Fig4}.

  The azimuthal velocity $v_{\phi}$ is symmetric with respect to $\theta=\pi/2$.
The direction of $v_{\phi} >0$ means that the advection velocity in eq.
(\ref{vsum.eqn}) reduces.
That is, the Hall drift-velocity is cancelled by the induced shear flow.
The magnitude of $v_{\phi}$ is of the same order as $|v_\theta|$.
Its maximum is $v_{\phi}\approx 4$ cm yr$^{-1}$
at $\theta\approx  0.2\pi, 0.8\pi$, and $r\approx 0.98R$,
as shown in the right panel of Fig.~\ref{Fig4}.
As \citet{2019MNRAS.486.4130L} discussed
with respect to a two-dimensional plane-parallel simulation,
the plastic flow is important in the low-viscosity region,
where $\nu=10^{36}$-$10^{37}$ g cm$^{-1}$ s$^{-1}$.
The viscosity coefficient (eq.(\ref{nuvis_express}))
in our model decreases in the radial direction: 
$\nu=10^{37}$g cm$^{-1}$ s$^{-1}$ at $r=0.95R$
and $\nu=10^{36}$g cm$^{-1}$ s$^{-1}$ at $r=0.97R$.
Thus, high-velocity regions in the vicinity of the surface
correspond to low-viscosity regions.
As the viscosity coefficient decreases,
the spatial region relevant to $\nu<10^{37}$g cm$^{-1}$ s$^{-1}$
extends inwards, and the motion becomes global.

  Figure \ref{Fig4} presents a snapshot obtained at 
$0.02\tau_{\rm H}\approx 1.3\times 10^4$yr, 
but the overall structure is almost fixed with respect to time.
In addition, the spatial profile does not depend to a significant extant on 
the viscosity.  In contrast,
the magnitude of the velocity changes with the magnetic evolution and
substantially depends on the viscosity.
As estimated in eq.(\ref{ratiothtv}), 
the induced velocity is proportional to $\nu^{-1}$.
We found that the velocity is proportional to $\nu^{-1}$
in high-viscosity models ($\gg\nu_{c}$), and that
the scaling does not hold in our models with $0.2\nu_c$ and $\nu_c$.
That is, the low-viscosity models with $0.2\nu_c$ and $\nu_c$
correspond to a nonlinear regime, in which
the magnetic fields are also modified as a result of the back-reaction.

\subsection{Magnetic-field evolution with bulk flow}
\begin{figure}
\begin{center}
  \includegraphics[scale=1.0]{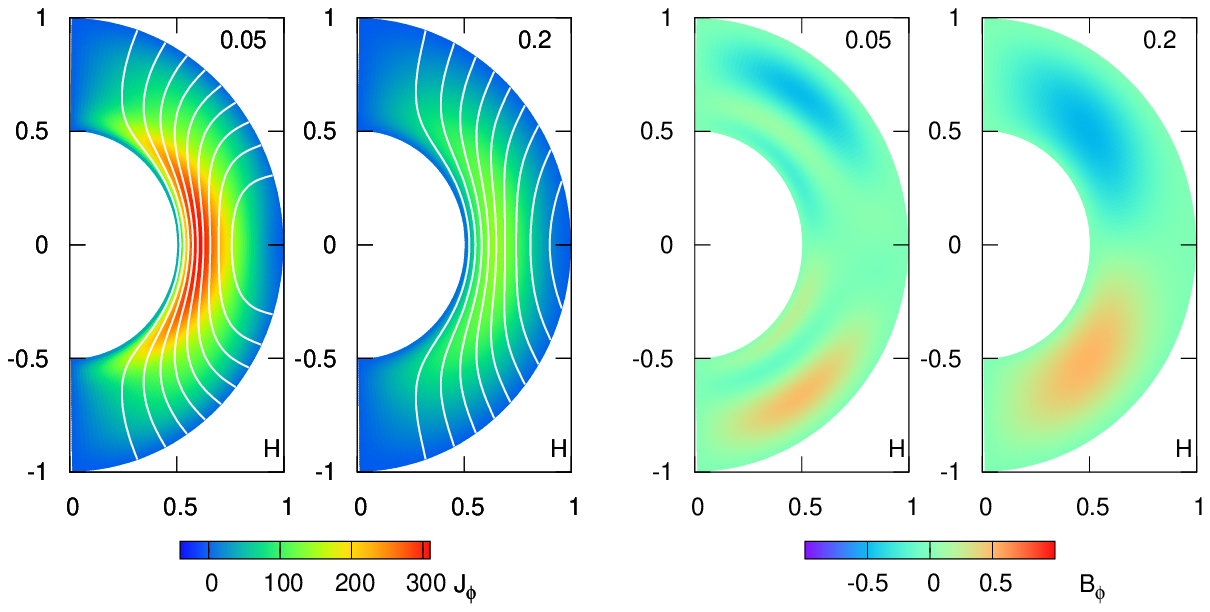}%
\caption{ 
\label{Fig5}
Magnetic field evolution 
from $0.05\tau_{\rm H}$ to $0.2\tau_{\rm H}$
in a high-viscosity model with $5\nu_{c}$. 
The two left panels show the 
magnetic function $\Psi$ as indicated by contour lines and 
azimuthal current $4\pi j_{\phi}R/(cB_0)$ indicated by colors 
in enlarged coordinates $(\xi(r)\sin\theta, \xi(r)\cos\theta)$. 
The two right panels show the evolution of the toroidal magnetic field 
$B_{\phi}/B_{0}$ as indicated by the color contour.
}
\end{center}
\end{figure}

In Fig.~\ref{Fig5}, the magnetic fields of a high-viscosity model 
with $5\nu_{c}$ are presented at two specific times: $0.05\tau_{\rm H}$ and at $0.2\tau_{\rm H}$.
They represent a rapidly decaying phase at the early time  
and a slowly decaying phase. 
After the time $0.2\tau_{\rm H}$, the time variation becomes 
much slower as inferred from the energy variation presented in Fig.~\ref{Fig3}.
The azimuthal current $j_{\phi}$ indicated by the color contours in the two left panels
does not change significantly in the distribution. 
The current is preserved near the core--crust boundary.
However, the magnitude decreases roughly by half at this interval.
The poloidal field lines indicated by the contours of $\Psi$ 
are slightly moved outwards, 
but the end points at the surface are almost fixed as the initial form.
That is, the external dipole field is unchanged.

The evolution of toroidal magnetic fields is presented in the two right panels
of Fig.~\ref{Fig5}.
The field $B_{\phi}$ is set as zero as our initial condition, 
but it is produced by the Hall drift (see eq.(\ref{initialevl.eqn})).
The spatial pattern depends on the initial dipolar
profile ($\propto\nabla _{\theta}\Psi\propto\cos\theta$) 
and electron fraction ($\propto\nabla _{r}\chi$).
Therefore, the toroidal field is anti-symmetric with respect to $\theta=\pi/2$.
The strength of the induced toroidal fields
is of the order $|B_{\phi}|\sim B_{0}$,
where $B_{0}$ is a magnitude of the dipolar polar cap at the surface. 
This value is small, as the maximum of $B_{r}$ is $\sim B_{0}$ and 
that of $B_{\theta}$ is $\sim 10B_{0}$ in the crust.
$|B_{\phi}|\sim B_{0}$ is understood  by observing eq.(\ref{initialevl.eqn}):
the toroidal field $B_{\phi}$ is related to a smaller
component $B_{r}(\propto\nabla_{\theta}\Psi)$ and $\nabla_{r}\chi$, 
which is also small owing to an almost flat electron-fraction distribution.

\begin{figure}
\begin{center}
  \includegraphics[scale=1.0]{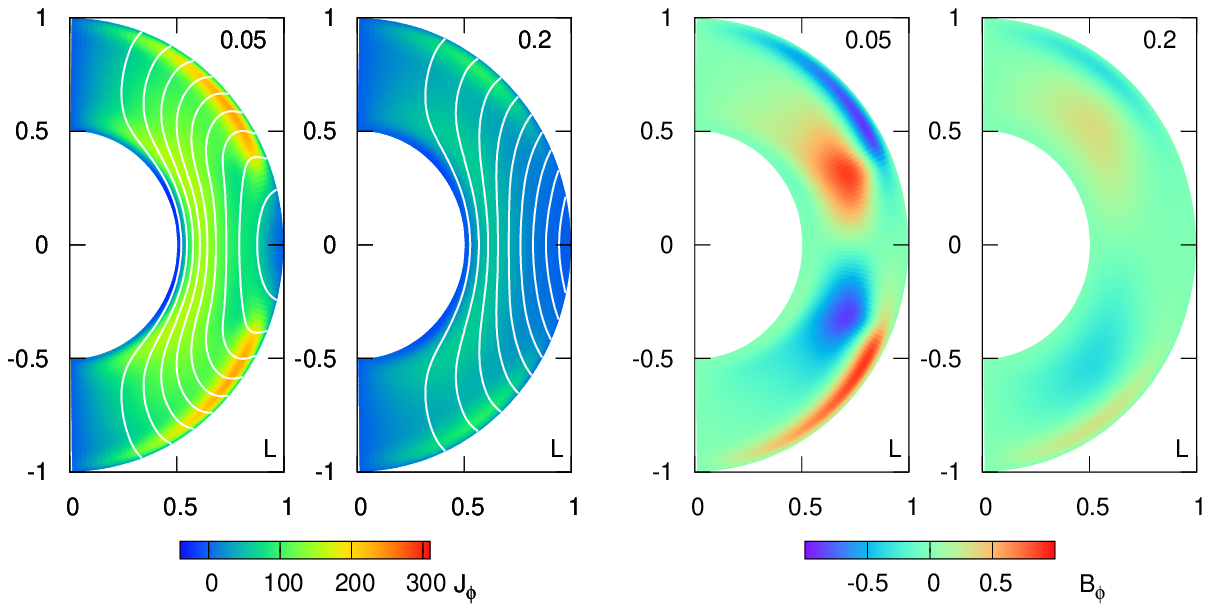}
\caption{ 
\label{Fig6}
Magnetic field evolution 
from $0.05\tau_{\rm H}$ to $0.2\tau_{\rm H}$
in a low-viscosity model with $0.2\nu_{c}$.
The same color-scale as that in Fig.~\ref{Fig5} is used
to compare the two models.
}
\end{center}
\end{figure}

The results for a low-viscosity model with $0.2\nu _{c}$
are presented in Fig.~\ref{Fig6}.
As shown in the two left panels, the distribution of $j_{\phi}$ 
is changed from the initial model.
In addition to a peak caused by the initial 
condition near the core--crust boundary, two strong current regions 
appear in the vicinity of the surface with a middle latitude:
$0.2\pi<\theta<0.4\pi$ and $0.4\pi<\theta<0.8\pi$.
This new distribution affects the external magnetic field.
We found that the field is almost described by dipole ($l=1$)
and octa-pole $(l=3)$ components, although  the latter is  
approximately one order of magnitude smaller than that of the dipole.

The toroidal magnetic field in a low-viscosity model with $0.2\nu_{c}$
is shown in the two right panels of Fig.~\ref{Fig6}.
On comparing it with Fig.~\ref{Fig5},
it is clear that the field strength of $B_{\phi}$ 
increases on reducing the viscosity coefficient.
The overall viscosity coefficient differs by a factor 25, but
the maximum of $|B_{\phi}|$ in Fig.~\ref{Fig6}
increases by a factor of a few than that in Fig.~\ref{Fig5}.
The magnetic-field dissipation is also enhanced, such that
high-velocity fluid motion does not effectively result in a stronger 
toroidal field.
It is also interesting to compare the low-viscosity model 
with the high-viscosity model in a late-time spatial pattern. 
In Fig.~\ref{Fig5}, we observe 
$B_{\phi}<0$ for the majority of the upper region,
whereas $B_{\phi}>0$ for the majority of the 
upper region in Fig.~\ref{Fig6}.
The directions of $B_{\phi}$ at $0.2\tau_{\rm H}$
in the low- and high-viscosity models are opposite.

One might be confused by the dependency of our results on the viscosity $\nu$. 
An inviscid fluid corresponds to $\nu =0$ and we have the Hall evolution 
in that case, as shown in subsection 3.3.
However, we here show that a smaller value of $\nu$ effectively changes 
from the magnetic field evolution obtained without considering the bulk motion.
In order to elucidate this fact, we first consider a model in mechanics:
a particle is falling  along a vertical axis under gravity and a drag force.
Equation of motion is given by
$dv/dt = -\nu v -g $, where $\nu $ and $g$ are positive constants.
By integrating this equation, we find that the velocity $v$ 
approaches a constant
$ v_{*} \equiv -g \nu^{-1}$ for $t \gg t_{*} \equiv \nu^{-1}$.
The terminal velocity $v_{*} $ satisfies $dv/dt =0$.
As the drag coefficient $\nu$ decreases, $v_{*} $ increases.
The timescale $t_{*}$ also increases.
In an extreme case $t_{*}$ may exceed the timescale concerned 
and approximation of $v= v_{*} $ breaks down.
Thus, we should be careful about the inviscid limit, $\nu=0$.
Our quasi-stationary assumption corresponds to $dv/dt =0$(see section 2.1) 
and the bulk velocity is determined by the condition.
In the high-viscosity model, the resultant velocity is too small to change magnetic field evolution.
That is, the results are the same as those obtained without bulk velocity.
As the viscosity $\nu$ decreases, the velocity increases and differences in magnetic fields become prominent. 
For example, toroidal field strength increases with the decrease of $\nu$.
Thus, the high-viscosity model with $5 \nu_{c}$ is the same as the Hall evolution, whereas the low-viscosity model with $0.2 \nu_{c}$ is quite different.
Like an illustrated model, it is impossible to obtain inviscid results by taking the limit of $\nu=0$, since our quasi-stationary assumption is no longer valid.
The timescale $\tau_{\rm v}$ (eq.(\ref{timescale.tv})) becomes too short.
%

 \section{Discussion}
  In this study, we considered
the effects of bulk motion on the magnetic-field evolution in a neutron-star crust.
The electric currents were settled to the barotropic MHD equilibrium 
state\citep{2013MNRAS.434.2480G},
and ions are locked at the lattices in a short timescale
($\ll 1$yr) after a magnetized star is newly born.
As the magnetic field is evolved in a secular timescale ($>10^3$yr), 
the Lorentz force also deviates from the initial form.
The magnetic stress in the crust increases simultaneously.
The response of the lattice ions is initially elastic, but plastic 
flow arises beyond the elastic yield limit.
In this study, the global circulation driven by the Lorentz force is simulated based
on the assumption that the circulation motion occurs at all times in the entire crust.

A large amount of electric currents is redistributed by the Hall drift, and
the magnetic energy is typically dissipated by
approximately $ 10^5(B_{0}/10^{13.5}{\rm G})^{-1}$yr,
where $B_{0}$ is the surface dipole field-strength at the polar cap. 
By taking into consideration the bulk motion, the
magnetic energy converts to a new channel, i.e.,
the bulk flow energy, and the loss rate is therefore accelerated.
The fraction to the bulk motion is typically 30\% of 
the magnetic-energy loss,
and the remaining is converted into heat.
The amount of work loss depends on the magnitude of the viscosity, as
the motion is effectively driven in a low-viscosity region
with $\nu=10^{36}$-$10^{37}$ g cm$^{-1}$ s$^{-1}$.
Our results consistently confirm the results of a previous local 
simulation\citep{2019MNRAS.486.4130L}.
Qualitatively, the bulk velocity induced by the Lorentz force is 
important in low-viscosity and strong-magnetic-field regimes
(see eq.(\ref{ratiothtv})).
In the case of the total amount of work, the volume is also an important factor.
In addition to the increase in the velocity driven by the viscosity,
the relevant volume increases on reducing the overall viscosity coefficient. 
Thus, the bulk motion is significantly important as a whole. 
 We also found new features in magnetic fields driven by bulk motion.
A high-velocity flow in a low-viscosity model results in a different 
pattern of toroidal fields.
An additional octa-pole poloidal field is also induced.
The new components of the toroidal and poloidal fields 
are small. They are of the order of 0.1 of the initial dipole in magnitude.
The overall magnetic field structure is not significantly changed.

  It is true that the global patterns of the flow and magnetic fields
depend on the initial magnetic-field configuration.
Our model comprises a simple dipole without a toroidal component.
In realistic cases, the fields may not be so smooth and could contain
irregular higher multipoles.
The consequence of the local irregularities is very important.
They may affect the global dipole field structure
or they may be cleaned up by a dynamical transition such as bursts.
For this purpose, it is necessary to connect the crustal
evolution to a magnetosphere\citep{2018MNRAS.481.5331A}.
We demonstrated global circulation motions in the entire crust
under simplified assumptions, and it was found to be desirable to
examine them in more realistic cases in the future.

\section*{Acknowledgements}
We thank Yuri Lewin for helpful comments.
This work was supported by JSPS KAKENHI Grant Number JP19K03850.

 \bibliographystyle{mnras}
 \bibliography{kojima20Apr} 
\end{document}